\documentclass[prl,twocolumn,graphicx,amssymb,floatfix]{revtex4-1}
\usepackage{graphicx}
\usepackage{color}
\usepackage{soul}
\usepackage{float}
\begin{document}

\title{A Comment on ``Asking photons where they have been in plain language''}
%\centerline{}
%\centerline{}
\author{ Lev Vaidman}
\affiliation{  Raymond and Beverly Sackler School of Physics and Astronomy,
 Tel-Aviv University, Tel-Aviv 69978, Israel
}

\noindent

\begin{abstract}
The criticism of the experiment showing disconnected traces of photons passing through a nested Mach-Zehnder interferometer is shown to be unfounded.
\end{abstract}
\maketitle

In a recent Letter \cite{sokol} Sokolovsky analyzes the experiment by Danan {\it et al.} \cite{Danan} and the theoretical proposal for this experiment \cite{past}.  The experiment shows a disconnected trace left by the the photons in the interferometer. Sokolovski agrees that the calculations in the theoretical paper are correct and that the results of the experiment are expected. However, he argues that the conclusion of \cite{Danan} and \cite{past}  that ``the past of the photons is not represented by continuous trajectories'' is  unjustified. He writes:  ``A simple analysis by standard quantum mechanics shows that this claim is false.'' In this Comment I will clarify the meaning of the results of \cite{Danan} and \cite{past} and refute the criticism of Sokolovski.

First, standard quantum mechanics certainly cannot show that the discussed sentence is false. In the framework of standard quantum mechanics photons do not have trajectories of any type.
From the text and the references, I understand that  Sokolovski considers Feynman's paths formulation of quantum mechanics as ``standard quantum mechanics''. While it might be considered as a standard calculational tool, only small minority attach to Feynman's paths ontological meaning. Indeed, papers  \cite{Danan} and \cite{past} have no arguments against continuity of Feynman's paths. But these paths do not represent a useful picture of the past of the photon. Sokolovski analyzes the arms of the interferometer, but Feynman's paths of the photons are everywhere. Every continuous line between the source and the detector is a Feynman path of the photon. Independently of the design of the interferometer, the photons are everywhere. Not an interesting answer to the question  the photons were asked in  \cite{Danan}: ``Where have they been?''.

All what experiment \cite{Danan} shows is that there are cases in which the past of the photon, defined in \cite{past}  as places with a significant weak trace, have parts which are not connected by continuous lines to the source and the detector. And this is a nontrivial example, since in most experiments with interferometers all  traces are  connected.

To avoid misunderstanding, I  repeat that all the results of \cite{Danan} and \cite{past} can also be explained by standard quantum mechanics as it has already  been stated in these works. Also, ``a prudent advice for its authors to double check that in Fig. 2 of \cite{Danan} the signal at the frequencies $f_E$ and $f_F$ is indeed absent, and not just too small to be seen against the background noise'' is not needed.  A ``tiny leakage of light in the inner interferometer'', which leads to these signals below the noise, is explicitly mentioned in \cite{Danan} and calculated, (Eq.8), in \cite{past}. A special status of regions $E$ and $F$, where the trace is nonzero but negligible relative to the regions $A$,  $B$ and $C$, is discussed in \cite{trace}.

A representative example of the ``plain language'' of Sokolovski: ``The particle remains in a real pathway combining all interfering paths'' is not part of the language of standard quantum mechanics since its formalism has no concept for the past of a pre- and post-selected particle. The past defined as regions with a weak trace is also not part of the standard formalism. However, Sokolovski showed nothing ``false'' or inconsistent about this approach.

This work has been supported in part by the Israel Science Foundation Grant No. 1311/14,
the German-Israeli Foundation for Scientific Research and Development Grant No. I-1275-303.14.

\end{document}